# Unidirectional zero reflection as gauged parity-time symmetry

James Gear[1], Yong Sun[2], Shiyi Xiao[1,3], Liwen Zhang[2], Richard Fitzgerald[1], Stefan Rotter[4], Hong Chen[2*] and Jensen Li [1*]

[1] School of Physics and Astronomy, University of Birmingham, Birmingham B15 2TT, UK
[2] Key Laboratory of Advanced Micro-structure Materials (MOE) and School of Physics Sciences and Engineering, Tongji University, Shanghai 200092, China
[3] School of Communication and Information Engineering, Shanghai University, Shanghai 200072, China
[4] Institute for Theoretical Physics, Vienna University of Technology (TU Wien), A-1040 Vienna, Austria, EU

* hongchen@tongji.edu.cn, j.li@bham.ac.uk

We introduce here the concept of establishing Parity-time symmetry through a gauge transformation involving a shift of the mirror plane for the Parity operation. The corresponding unitary transformation on the system's constitutive matrix allows us to generate and explore a family of equivalent Parity-time symmetric systems. We further derive that unidirectional zero reflection can always be associated with a gauged PT-symmetry and demonstrate this experimentally using a microstrip transmission-line with magnetoelectric coupling. This study allows us to use bianisotropy as a simple route to realize and explore exceptional point behaviour of PT-symmetric or generally non-Hermitian systems.

**Contents**

**1. Introduction**

**2. Gauged PT-symmetry condition**

**3. Bianisotropic transmission line**

**4. Unidirectional zero reflection as passive PT-symmetry**

**5. Conclusion**

## 1. Introduction

Parity-time (*PT*) symmetric Hamiltonians have been proposed as a class of non-Hermitian Hamiltonians with real eigenvalues that could possibly generalize the conventional paradigm of quantum mechanics [1-3]. At the so-called *PT*-symmetry breaking transition, these real eigenvalues evolve into complex conjugate pairs - a phenomenon that has triggered a series of developments, particularly in optical analog studies, where power oscillations [4-7], coherent *PT*-symmetric laser/absorbers [8-10], optical solitons in *PT* periodic systems [11], rainbow nonreciprocity [12], and wave propagation with unidirectional zero reflection, unidirectional invisibility, or constant intensity propagation [13-17], have been demonstrated. From the technological point of view, these developments have turned into a unique approach to achieve tuneable components with extreme sensitivity and unconventional behaviours [18-20]. A *PT*-symmetric Hamiltonian can be realized by optical components with a balanced gain/loss pair [5-7,9-13]. This is understood as the ideal configuration of *PT*-symmetry. Further studies have shown that *PT*-symmetry breaking can also occur in passive systems, which can be mapped back to the ideal *PT* symmetric Hamiltonian by biasing the system with the averaged level of loss [4]. Such a gain/loss unbalanced situation is called a passive *PT* symmetry [4,14]. More recently, *PT*-symmetry has also been introduced into metamaterials [21-26]. As metamaterials provide very flexible constitutive parameters [27-29], one has the ability to realize more complex PT-symmetric systems in the subwavelength regime. Highlights here are the possibility of unidirectional cloaking, *PT*-symmetric transformation optics [21,22], exceptional point-enabled polarization manipulation [23,24], the realization of discrete breathers [25], and reversible nonreciprocity [26].

In this work, we generalize Parity-time symmetry by gauging the Parity operator through a continuous parameter, which represents the location of the mirroring plane. As we shall see, such a gauge generalization gives us a way to generate an equivalent family of Parity-time symmetric systems from a given seed system. It allows us to extend many of the special wave phenomena associated to Parity-time symmetry to a much broader and equivalent family. Here, we focus on a particular phenomenon called unidirectional zero reflection (UZR) at the exceptional point of a PT-symmetric system, with zero reflectance from one side, but not from the other side of incidence [13-15]. Interestingly, UZR may be able to occur in other conventional optical systems as well, for example systems with bianisotropy, in which the forward and backward impedances are different [30-32, 37,38]. Here we provide a common framework for understanding UZR in all of these different contexts by showing that UZR can actually always be interpreted as a result of passive PT-symmetry through a gauged parity operation. To establish this connection between UZR and passive PT-symmetry experimentally, we use a microstrip transmission-line with bianisotropy to verify the validity of our theoretical predictions.

## 2. Gauged PT-symmetry condition

We start from a transmission-line system with voltage $V$ and current $I$ for a generic discussion of one-dimensional wave propagation. Figure 1(a) shows the black-box representation of such a transmission-line system. The voltage at port 1 and port 2 are defined as $V_1$ and $V_2$ while the currents flowing to the right at the two ports are defined by $I_1$ and $I_2$ respectively. The complex reflection amplitude for forward (backward) incidence is $r_f$ ($r_b$), defined here at a common reference plane (the vertical black line), that is situated half-way in between the two ports.

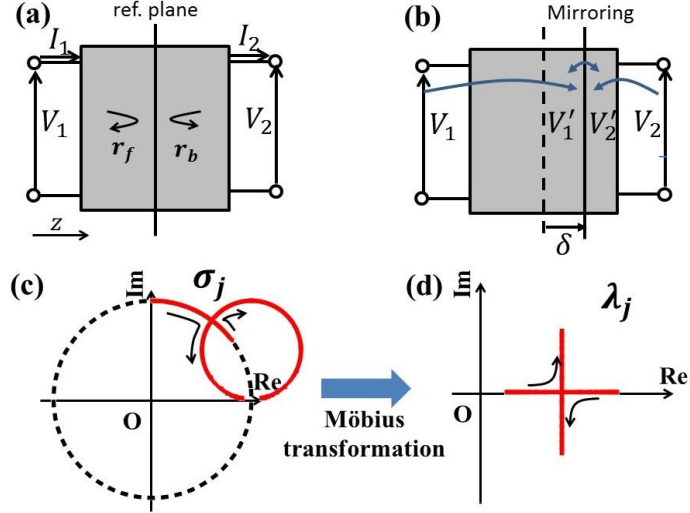

Figure 1. (a) Transmission-line black-box representing one-dimensional wave propagation along the $z$-direction. Voltage at port 1 (2) is represented by $V_1$ ($V_2$). Current at port 1 (2) is represented by $I_1$ ($I_2$) with positive value for current flowing to the right. The grey area denotes a black-box representation of the system. The forward (backward) complex reflection amplitude $r_f$ ($r_b$), is defined with a common reference plane in the middle of the system, the vertical black line. (b) The gauging action by shifting the reference plane by $\delta$ in the $z$-direction with all the relevant quantities now in primed notation, are defined at the shifted reference plane. (c) and (d) are the trajectories of eigenvalues $\sigma_j$ of the scattering matrix and $\lambda_j$ ($j \in 1,2$) of the constitutive matrix in the complex plane for an ideal PT-symmetric system, going from the PT-symmetric phase to the broken phase through a degeneracy of eigenvalues at the exceptional point.

To discuss PT-symmetry, we need to consider the action of parity and time-reversal operations on the various quantities. An obvious choice for the parity operation is the mirroring action with respect to the reference plane shown in figure 1(a), i.e. $V_1 \leftrightarrow V_2$, $I_1 \leftrightarrow -I_2$. However, by using the above black-box description, one can immediately recognize that there is a degree of freedom in where we put such a mirror plane. This degree of freedom becomes more apparent when we discuss metamaterial structures having both electric and magnetic responses in the next section as there is no obvious way to associate a mirror action plane. In light of this, figure 1(b) shows the situation if we put our mirror action plane at a location shifted from the reference plane by a distance $\delta$ in the forward propagation direction. This shifted mirroring action essentially means that we shift our reference plane for the voltages and the currents to a new reference plane (as if they are propagating on the background transmission line), then perform the mirroring action by swapping the voltages and currents on the two sides, and finally shift the reference planes back to the original position. The variable $\delta$ now acts like a gauging parameter to choose the definition of the P-operator, which can now be expressed as

$$P \begin{pmatrix} V_j \\ I_j \end{pmatrix} = \begin{pmatrix} \cos(2k\delta) & -i\sin(2k\delta) \\ i\sin(2k\delta) & -\cos(2k\delta) \end{pmatrix} \begin{pmatrix} V_{3-j} \\ I_{3-j} \end{pmatrix}, \qquad (1)$$

where $k$ is the wave-number of free-space (the background transmission line, in this case). This P-operator has many of the properties of the usual parity operator (such as $P^2 = I$, being unitary and Hermitian) so we should be able to pair it with the time-reversal operator

$$T \begin{pmatrix} V_j \\ I_j \end{pmatrix} = \begin{pmatrix} V_j^* \\ -I_j^* \end{pmatrix}, \qquad (2)$$

to derive a combined PT-symmetry condition with the fore-mentioned gauging parameter in the next step. On the other hand, the voltage $V$ and current $I$ on the two sides are related to each other through a constitutive matrix description:

$$\begin{pmatrix} I_2 - I_1 \\ V_2 - V_1 \end{pmatrix} = ikL \begin{pmatrix} 1 + \chi_{e,\text{eff}} & \xi_{\text{eff}} \\ -\xi_{\text{eff}} & 1 + \chi_{m,\text{eff}} \end{pmatrix} \cdot \begin{pmatrix} (V_2 + V_1)/2 \\ (I_2 + I_1)/2 \end{pmatrix}$$
$$= i \left( kL \begin{pmatrix} 1 & 0 \\ 0 & 1 \end{pmatrix} + Y \right) \begin{pmatrix} (V_2 + V_1)/2 \\ (I_2 + I_1)/2 \end{pmatrix}. \quad (3)$$

where $L$ is the physical thickness of the system. We label the matrix elements of the constitutive matrix $Y$ as $kL\chi_{e,\text{eff}}$, $kL\xi_{\text{eff}}$, and $kL\chi_{m,\text{eff}}$. They represent the surface susceptibilities for the case of electromagnetic waves. We also limit ourselves to reciprocal systems in this work so that the two off-diagonal elements of the Y-matrix have same value but opposite signs. Nevertheless, Eq. (3), relating the difference of fields to the averaged fields, is the wave equation in the continuum limit except that we now model the system as a black-box. If the system is PT-symmetric in its response then Eq. (3) should be invariant after applying the combined PT-operation: $[PT, Y] = 0$. To begin with a "conventional" case, we set $\delta = 0$ (no gauging) so that the parity operator is the usual mirror operation along the propagation direction, written as $M = P(\delta = 0)$. Invariance under this operation immediately induces the PT-symmetry condition (using $[PT, Y] = 0$) to be the requirement that the effective constitutive parameters of matrix $Y$ are all real numbers. By having non-zero off-diagonal elements, we can hit the exceptional point where the two eigenvalues and the two eigenvectors of the Y matrix coalesce. Such a PT-symmetric constitutive matrix can be realized by stacking two dielectric slabs with the real parts of their permittivities being the same but their imaginary parts having opposite signs (e.g. $\chi_{e,1} = a + i\gamma$ and $\chi_{e,2} = a - i\gamma$). Equivalent behaviour can also be achieved by a pair of metasurfaces with shunt resistances of the same magnitude but opposite signs (gain and loss) [43]. Now, with the introduced gauging parameter $\delta$, we can define other kinds of PT-symmetry. As another typical case, we put $k\delta = \pi/4$ for which we obtain the following PT-symmetry condition:

$$\chi_{e,\text{eff}} = (\chi_{m,\text{eff}})^* \text{ and } \xi_{\text{eff}} = -(\xi_{\text{eff}})^* \quad (4)$$

To realize this, one can again consider a pair of slabs stacked together. One slab is purely dielectric with $\chi_{e,1} = a + i\gamma$ (real part $a$ and imaginary part $\gamma$), $\chi_{m,1} = 0$ while another slab is purely magnetic $\chi_{m,2} = a - i\gamma$, $\chi_{e,2} = 0$. This is a special case of our gauged PT-symmetry, as previously found by matching permittivity and permeability in Ref. [44]. While this double-slab system does not seem to have an obvious mirror symmetry even if we turn off the gain/loss by setting $\gamma \to 0$, it can now be considered as having PT-symmetry with a particular gauge $k\delta = \pi/4$ by setting a particular location of the mirror plane. From this perspective, the gauging parameter $\delta$ actually introduces a whole family of PT-symmetric systems. For the general case of arbitrary $\delta$, the family of PT-symmetric constitutive matrices have their family members related to each other simply through a unitary transformation:

$$\begin{pmatrix} \chi'_{e,\text{eff}} & \xi'_{\text{eff}} \\ -\xi'_{\text{eff}} & \chi'_{m,\text{eff}} \end{pmatrix}_{PT} = U \begin{pmatrix} \chi_{e,\text{eff}} & \xi_{\text{eff}} \\ -\xi_{\text{eff}} & \chi_{m,\text{eff}} \end{pmatrix}_{PT} U^\dagger \quad (5)$$

where

$$U = \begin{pmatrix} \cos(k\delta) & i\sin(k\delta) \\ i\sin(k\delta) & \cos(k\delta) \end{pmatrix}.$$

Eq. (5) can be regarded as the gauge transformation subject to the gauge parameter $\delta$. It means that any constitutive matrix obtained from a seed PT-symmetric constitutive matrix through a unitary transformation (the defined $U$) can be equivalently regarded as PT-symmetric, with the parity operator being gauged to another mirror action plane by a shift of distance $\delta$. We emphasize here that we put our focus on the PT-symmetry condition for the system response, i.e., on the constitutive matrix directly. This is relevant since systems whose internal system structure is not necessarily PT-symmetric (e.g. the systems discussed in [17]), may satisfy the PT-symmetry condition in its constitutive matrix. Consider here as an example the case of $k\delta = \pi/4$, for which the matrix $Y$ becomes

$$Y = \begin{pmatrix} a + i\gamma & i\kappa \\ -i\kappa & a - i\gamma \end{pmatrix} \tag{6}$$

with $a$, $\gamma$ and $\kappa$ being all real numbers. Interestingly, this Y-matrix has the same form as a 2×2 PT-symmetric Hamiltonian in quantum mechanical systems [1,2], which is Hermitian for the case without loss and gain. One can also derive the scattering matrix S from Eq. (3), with convention $S = \{\{t, r_b\}, \{r_f, t\}\}$. The Y and S matrices are then simply related to each other through a Möbius transformation (matrix version):

$$Y = \frac{2}{i} B \frac{S - I}{S + I} B^{-1} \text{ with } B = \begin{pmatrix} 1 & 1 \\ 1 & -1 \end{pmatrix}, \tag{7}$$

with $I$ being the 2×2 identity matrix [49]. Here, the S-parameters are defined on the common reference plane (figure 1(a), or a zero thickness $L$ for the system) and the $Y$ matrix obtained represents the surface susceptibilities mentioned earlier. Now, taking the example of the Y matrix in Eq. (6), as we vary the value of the gain/loss parameter $\gamma$, we can plot the eigenvalues ($\sigma_1$ and $\sigma_2$) of the scattering matrix, as shown in figure 1(c), in the complex plane. When $\gamma$ increases from zero in the PT-symmetric phase, $\sigma_1$ and $\sigma_2$ trace out the unit circle in the complex plane, until a certain threshold when $|\gamma| = |\kappa|$, where we have an exceptional point. In the PT-broken phase, the eigenvalues separate and deviate from the unit circle with reciprocal moduli. At the exceptional point, where eigenvalues of $S$ become degenerate, UZR is realized with either $r_f$ or $r_b$ going to zero. In such a case, the complex transmission coefficient $t$ has unit magnitude [13].

On the other hand, the PT-symmetric behaviour can also be revealed if we look at the eigenvalues of the constitutive Y-matrix. The eigenvalues of the Y-matrix ($\lambda_1$ and $\lambda_2$) are related to those of the S-matrix by

$$\lambda_j = \frac{2}{i} \frac{\sigma_j - 1}{\sigma_j + 1}$$

where $j = 1,2$. They are plotted in figure 1(d). These eigenvalues appear either in pairs of real numbers (*PT*-symmetric phase) or of two conjugate complex numbers (*PT*-broken phase) correspondingly. We note that this behaviour resembles that of the eigenvalues of a PT-symmetric Hamiltonian and we therefore prefer the usage of the constitutive matrix instead of the S-matrix for representation. Interestingly, Eq. (6) has the same form of a 2×2 PT-symmetric Hamiltonian as that of a PT-symmetric optical waveguides pair or quantum-mechanical pair potential [2, 4, 12].

## 3. Bianisotropic transmission line

The above discussion on generalizing the PT-symmetry condition implies that many previously considered exotic phenomena related to PT-symmetry can be immediately extended to a generalized but equivalent family. We have also proposed that PT-symmetry can be conveniently described by the constitutive matrix, which plays a similar role to that of the Hamiltonian in discussing PT-symmetry in quantum mechanics. More specifically, for the case of UZR, by finding the constitutive matrix of the system, we will show that we can always associate such a system with passive PT-symmetry.

Next, we provide experimental support for validating the derived general PT-symmetry. For this purpose, we consider an electric resonating atom (a vertical bar with its far-end grounded) and a magnetic resonating atom (a split-ring resonator) coupled to a transmission line at various positions. Figure 2(a) shows a photograph of the sample when they are put at the same location. A natural choice for both the mirror and reference plane of PT-symmetry is indicated by the vertical black line. Figure 2(b) shows the case when the two atoms are put at different locations separated by a distance $\delta$, chosen as 20mm, while the extent of the losses at the two atoms can be varied by the two variable resistors mounted on the atoms. The two-ports are connected to a vector network-analyzer (Agilent N5232A) for measuring the scattering parameters. For example, when we set the variable resistor at the split-ring and the vertical bar at 20Ω and 0Ω, we can extract the constitutive matrix (Y) from the experimentally obtained scattering parameters using Eq. (7). When we have metamaterial structures, it is not straightforward to identify where the interface (between freespace and metamaterial) lies. Therefore, the S-parameters here are simply defined on a common reference plane at the position of the electric atom, effectively with a zero thickness of the sample. This means that when we extract the Y matrix, we are only including the activity of the resonating elements and not the background medium. In the graphs, we just label the matrix elements of $Y$ as $\{\{\chi_e,\xi\},\{\zeta,\chi_m\}\}$ for simplicity. The four matrix elements are plotted in figure 2(c), the electric response $\chi_e$ constitutes a Drude-type resonance at zero frequency for the electric atom with some additional electric response from the magnetic atom. The magnetic response $\chi_m$ is a Lorentzian, coming from the magnetic resonance of the split-ring. The bianisotropy, from the near-field coupling between the two atoms, is revealed through the non-zero values of $\xi$ or $\zeta$. In fact, the experimental results can be accurately represented by the following dipolar model:

$$S - I = A^{\dagger} \begin{pmatrix} -\frac{2i}{\chi_{eA}} - 1 & -e^{ik\delta} & e^{ik\delta} \\ -e^{ik\delta} & -\frac{2i}{\chi_{eB}} - 1 & 0 \\ -e^{ik\delta} & 0 & -\frac{2i}{\chi_{mB}} - 1 \end{pmatrix}^{-1} A,$$

$$\text{with } A = \begin{pmatrix} 1 & 1 \\ e^{ik\delta} & e^{-ik\delta} \\ e^{ik\delta} & -e^{ik\delta} \end{pmatrix}$$

(8)

where $\chi_{eA}$ is the electric response of the electric atom, $\chi_{eB}$ and $\chi_{mB}$ are the electric and magnetic response of the magnetic atom. The matrix $A$ converts incident fields to local fields. The matrix in the middle is the polarizability matrix to turn local fields to dipole moments while the matrix $A^{\dagger}$ converts the dipole moments to scattered waves. The phase factor $e^{ik\delta}$ at various positions means the coupling between the electric and magnetic atoms, where $k$ is the wavenumber of the background transmission line and $\delta$ is the distance between the two resonators. By using a Lorentz model fitting on the $\chi_{eA}$, $\chi_{eB}$ and $\chi_{mB}$ (see Appendix A for details), one can then obtain the Lorentz-model representation of the final structure and the corresponding Y-matrix elements are plotted as dashed lines in figure 2(c). They fit well with the measured results for the sample in figure 2(b) within the frequency range we are interested in (from 0.4 to 1GHz) and also reveal the bianisotropy from the shifting.

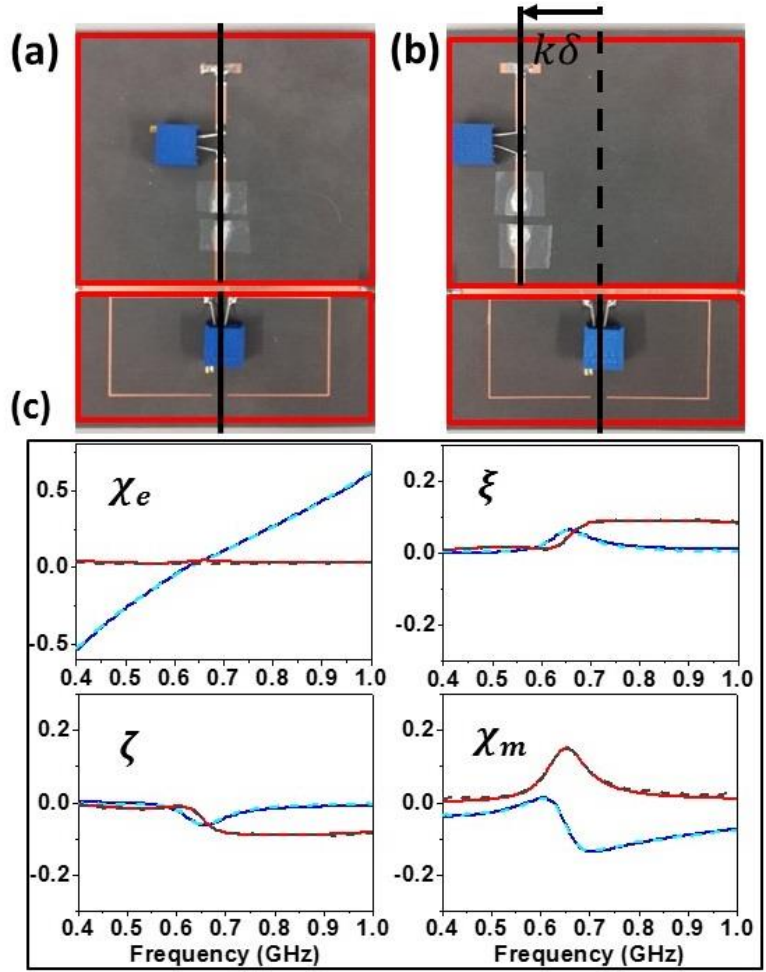

Figure 2. (a) Photo of a symmetric structure where the reference planes of the electric (upper red box) and magnetic (lower red box) resonators are aligned to the same position. (b) Photo of the structure with bianisotropy with variable resistance at the electric (magnetic) atom being set as 0 (20) Ω. (c) Y-matrix elements of the bianisotropic structure with reference plane at the electric atom. The solid blue and red lines denote the real and imaginary parts of the matrix elements extracted from measurements, respectively. The light blue and dark red dashed lines are the real and imaginary parts of the corresponding matrix elements from the theoretical dipolar model, respectively.

Now, suppose we increase the tuneable resistance R at the electric atom while keeping the resistance of the magnetic atom set at 20Ω. The constitutive Y-matrix is extracted again from the experimental results of the scattering parameters. They are shown in figure 3 for both the real and imaginary parts of the 4 matrix elements. One can see that R mostly affects $\chi_e$, with larger imaginary part generally at larger R. It also affects the real part of $\chi_e$ due to some induction effect (a coil structure is present within the variable resistor). The dipolar model for each value of R is fitted and serves as the model representation later.

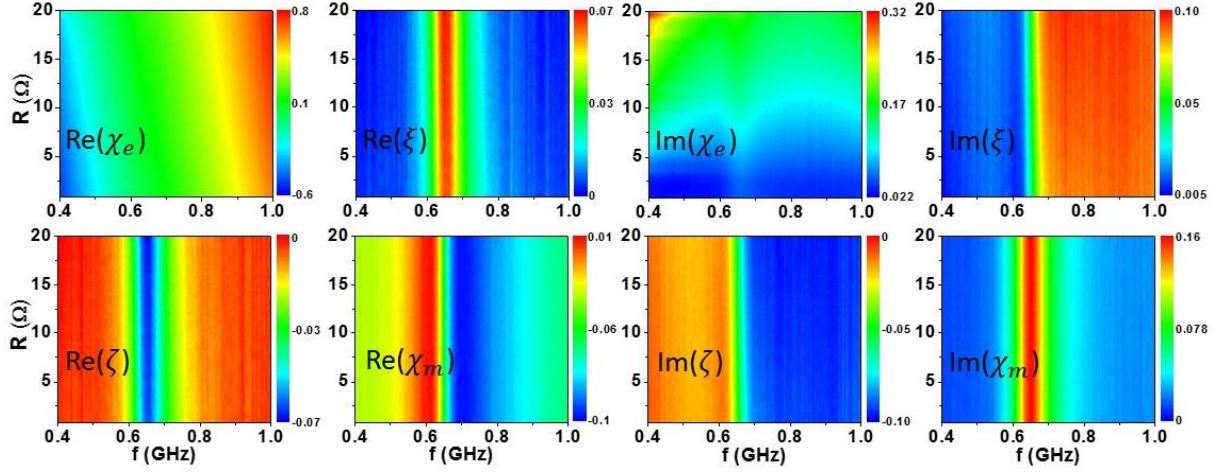

Figure 3. The real (left four panels) and imaginary (right four panels) parts of the experimentally extracted constitutive matrix elements as a function of the electric atom resistance (R) and the frequency in GHz (f). Reference plane is chosen to be at the electric atom.

By tuning the variable resistor of the electric atom, the bianisotropic transmission line can display UZR. We first locate where UZR can be obtained in the frequency-resistance phase space. The colour map in figure 4(a) shows a measure of UZR defined by $(|r_f| - |r_b|)/(|r_f| + |r_b|)$. It gives a value $-1$ for UZR at the first point (labeld by Ex1) with $(f_{Ex1}, R_{Ex1}) \cong (0.6 GHz, 3\Omega)$ and a value $+1$ for UZR at the point (labeled by Ex2) with $(f_{Ex2}, R_{Ex2}) \cong (0.62 GHz, 14\Omega)$. These two points are actually the exceptional points, i.e., the eigenvalue degeneracy of the scattering S-matrix. For example, we have plotted the trajectories of the eigenvalues of the S-matrix when the tuneable resistance is varied at fixed frequency $f_{Ex1}$, shown as blue colour in figure 4(b) (solid line for dipolar model results and symbols for the experimental results). The two eigenvalues cross each other and become degenerate at the exceptional point, at the tunable resistance $3\Omega$. Similarly, we can vary the tunable resistance at $f_{EX2}$ (plotted in red colour) to pass through the exceptional point with eigenvalue degeneracy and UZR. The UZR at exceptional point Ex1 has $r_b = 0$ and $r_f = 0$ at the second exceptional point Ex2. As we have discussed, an eigenvalue degeneracy in the S matrix can be translated to an eigenvalue degeneracy in the Y-matrix. We have plotted the eigenvalue trajectories in figure 4(c) with corresponding phenomena observed. The condition of the passive PT-symmetry, being formulated later, is plotted in advance as dashed line in figure 4(a) and it passes through the point of UZR as predicted.

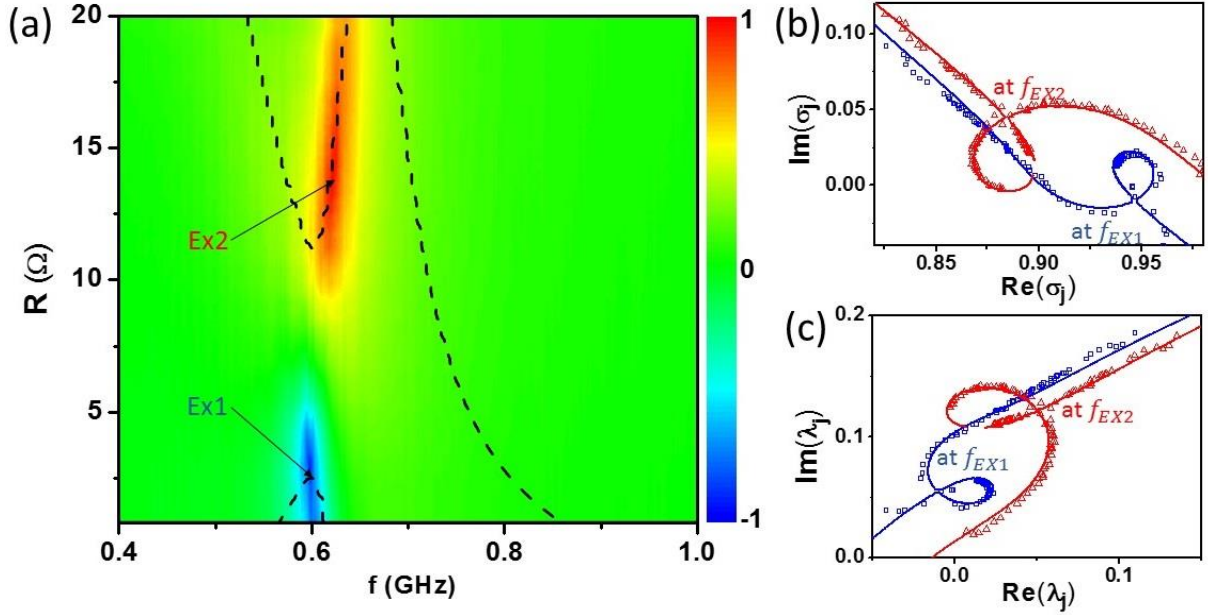

Figure 4. (a) Measure for UZR defined by $(|r_f| - |r_b|)/(|r_f| + |r_b|)$, extracted from experimental data. The dashed line shows where the system will be proved to be passive PT-symmetric in the next step. (b) Eigenvalues of S at the frequency of exceptional point 1 (blue) and exceptional point 2 (red) as a function of the tuneable resistance R. (c) Eigenvalues of Y at the frequency of exceptional point 1 (blue) and exceptional point 2 (red) as a function of the tuneable resistance R.

## 4. Unidirectional zero reflection as passive PT-symmetry

Up to now, we know that the exceptional point (eigenvalue degeneracy) of the Y-matrix is equivalent to the eigenvalue degeneracy for the S-matrix where we obtain UZR. It is well known that a conventional PT-symmetric system gives rise to UZR at an exceptional point but the converse is not a common statement, i.e. UZR does not guarantee that the system is PT-symmetric in the normal sense where P is a simple mirroring action. Here we generalize the connection between UZR and PT-symmetry by using the gauged parity operator in the following. In fact, UZR, an eigenvalue degeneracy of the Y-matrix, can be stated as $\chi_{e,eff} - \chi_{m,eff} = \pm 2\xi_{eff} = \mp 2\zeta_{eff}$ (reciprocity assumed so that $\xi_{eff} = -\zeta_{eff}$ is always satisfied). As the eigenvalue degeneracy of the Y-matrix retains even if we shift to another reference plane (as it is just a unitary transformation), we can always choose the gauge such that $Re(\chi_{e,eff}) = Re(\chi_{m,eff})$. In such a gauge, UZR means $Re(\xi_{eff}) = Re(\zeta_{eff}) = 0$ together with $Im(\chi_{e,eff} - \chi_{m,eff}) = \pm 2Im(\xi_{eff}) = \mp 2Im(\zeta_{eff})$. Now, we compare this to the form shown in Eq. (6). For a reciprocal system displaying UZR, we can therefore always write the Y-matrix for such a system as

$$Y = Y_{PT} + i\gamma_{bias}I, \qquad (9)$$

where $Y_{PT}$ is PT-symmetric (in the form shown in Eq. (6)). Hence, we can always associate a Y-matrix possessing UZR to a PT-symmetric one by biasing the diagonal values with a common imaginary part $\gamma_{bias}$. This can be called a passive PT-symmetric system, inspired by a similar nomenclature for considering PT-symmetry from a pair of optical waveguides with loss contrast [4]. Thus, we have generalized the notion of passive PT-symmetry when the Parity operator is now generalized to its gauged form. We emphasize that with such a generalization, Eq. (9) is the necessary and sufficient condition for obtaining UZR in a reciprocal system.

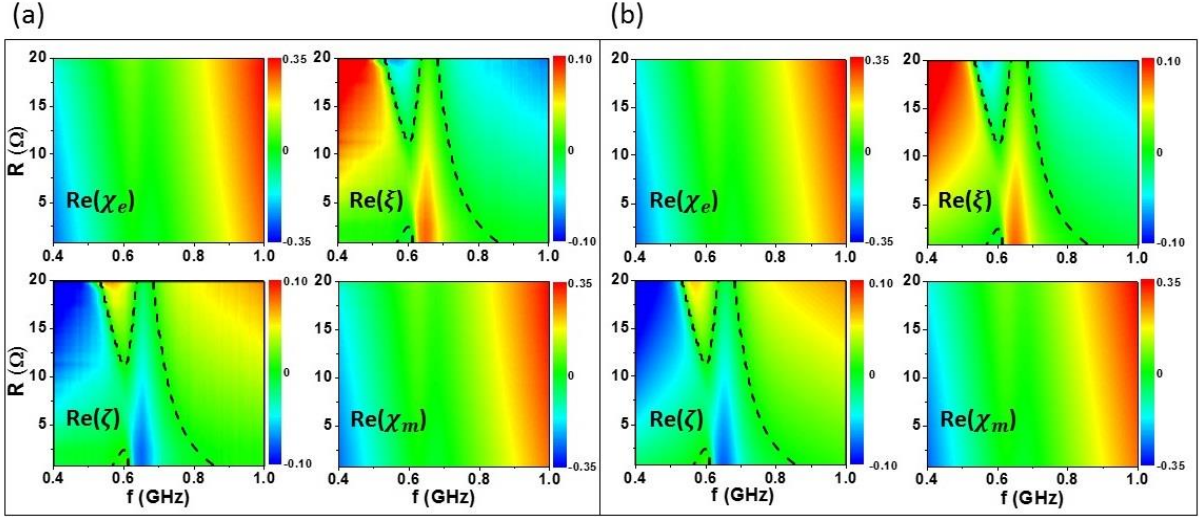

Figure 5. The constitutive matrix after choosing a gauge to have $Re(\chi_e) = Re(\chi_m)$, as a function of the electric atom resistance and the frequency $f$ in GHz. The four panels in (a) show the real part of Y-matrix elements extracted from experimental results. The four panels in (b) show the corresponding model results. The passive PT-symmetry condition is indicated by the dashed line where $Re(\xi) = Re(\zeta) = 0$.

For the case of our bianisotropic transmission line, the real part of the Y-matrix elements are plotted in figure 5 in the range of frequency and tunable resistance we are considering. As we have discussed, the Y-matrix is now gauged to the reference plane with the real part of $\chi_e$ and $\chi_m$ being the same, as shown in figure 5 (a). Figure 5(b) shows the corresponding dipolar model results with very good agreement. From the panel showing $Re(\xi)$ or $Re(\zeta)$, the contour with zero value is identified as the dashed-lines, these lines are where the system can be regarded as passive PT-symmetric. These lines pass through the exceptional points, where also the following identities hold: $Im(\chi_{e,eff} - \chi_{m,eff}) = \pm 2Im(\xi_{\text{eff}}) = \mp 2Im(\zeta_{\text{eff}})$ as figure 4 shows.

## 6. Conclusion

It has been known that unidirectional zero reflection (UZR) can occur in systems both with and without PT-symmetry, but the exact relationship between these two phenomena was not clear. By considering an extension of the Parity operator through gauging, we have established that a reciprocal system with UZR is always equivalent to a passive PT-symmetric system through a unitary transformation of the constitutive matrix, which effectively means a change of reference plane for the Parity operator. This link will be helpful to explore systems with hidden PT-symmetry, such as the bianisotropic transmission line considered as an example in this work. Moreover, the considered gauge transformation generalizes exceptional point wave phenomena associated with PT-symmetry to a broader family of equivalent systems containing exceptional points.

## Appendix A: Dipolar model

To model the bianisotropic structure we need to know the responses of the individual atoms. We describe the electric atom (atom A) and the magnetic atom (atom B) by the constitutive matrices $Y_A = \{\{\chi_{eA}, 0\}, \{0, 0\}\}$ and $Y_B = \{\{\chi_{eB}, 0\}, \{0, \chi_{mB}\}\}$, where we have assumed that $\chi_{mA}$ is negligible. We can then relate the susceptibilities of the A and B atoms to their polarizabilities via

$$1/\alpha_{eA} + i = 2/\chi_{eA},$$
$$1/\alpha_{eB} + i = 2/\chi_{eB},$$
$$1/\alpha_{mB} + i = 2/\chi_{mB}.$$

The additional imaginary constant $i$ is the so-called radiative correction at finite frequency. The $\chi_{eA}$, $\chi_{eB}$ and $\chi_{mB}$ can be extracted numerically through Eq. (7) from the S-parameters of the individual atoms. The extracted $\chi_{eA}$, $\chi_{eB}$ and $\chi_{mB}$ can be approximately decomposed into resonance terms

$$\chi_{eA} = f\left(a_{e0} + i\gamma_{e0} - \frac{a_{e1}}{f^2 + i\gamma_{e1}f} - \frac{a_{e2}}{f^2 - f_{e2}^2 + i\gamma_{e2}f}\right)$$

$$\chi_{eB} = f(b_{e0} + b_{e2}f^2)$$

$$\chi_{mB} = f\left(b_{m0} + i\gamma_{m0} + b_{m2}f^2 - \frac{b_{m1}f^2}{f^2 - f_{m1}^2 + i\gamma_{m1}f}\right),$$

where $f$ is the frequency in units of GHz. They are proportional to $f$ due to the definition in Eq. (3). The constant terms and the quadratic terms are the low-frequency Taylor series expansion contributed by the higher order modes that are not captured by the resonance terms. The various constants are extracted as $a_{e0} = 0.49$, $\gamma_{e0} = 0.0042R$, $a_{e1} = 0.36$, $\gamma_{e1} = 0.013 + 0.006R$, $f_{e2} = 1.57 - 0.0024R$, $a_{e2} = 0.83 + 0.0038R$, $\gamma_{e2} = 0.041 + 0.0016R$, $b_{e0} = 0.093$, $b_{e2} = -0.081$ and $b_{m0} = -0.15$, $\gamma_{m0} = 0.0037$, $f_{m1} = 0.65$, $b_{m1} = 0.041$, $\gamma_{m1} = 0.10$, $b_{m2} = 0.11$. These parameters serve as the starting parameters to fit the Y-matrix of the final structure with a shifted distance $\delta = 20 mm$ between the two atoms through the interaction dipolar model in Eq. (8).

## Acknowledgements

J.L. would like to acknowledge funding support from the European Union's Seventh Framework Programme (FP7) under Grant Agreement No. 630979. S.R. is supported by the Austrian Science Fund (FWF) through project No. F49 (SFB NextLite).